\documentclass[aps,preprint,showpacs,showkeys,nofootinbib,superscriptaddress,amsmath,amssymb]{revtex4-2}

\usepackage[utf8]{inputenc}
\usepackage[T1]{fontenc}
\usepackage{amsmath,amssymb,amsfonts,bm}
\usepackage{graphicx}
\graphicspath{{./}{figures_fl_oscillators/}}
\usepackage{xcolor}
\usepackage{float}
\usepackage{hyperref}
\usepackage{enumitem}
\usepackage{orcidlink}

\hypersetup{
    colorlinks=true,
    linkcolor=blue,
    citecolor=blue,
    urlcolor=blue
}

\newcommand{\be}{\begin{equation}}
\newcommand{\ee}{\end{equation}}
\newcommand{\bea}{\begin{eqnarray}}
\newcommand{\eea}{\end{eqnarray}}

\newcommand{\vp}{\mathbf{p}}
\renewcommand{\vr}{\mathbf{r}}
\newcommand{\vn}{\mathbf{n}}
\newcommand{\vu}{\mathbf{u}}

\newcommand{\Om}{\Omega}
\newcommand{\mapp}{m_{\mathrm{app}}}
\newcommand{\D}{\mathcal{D}}
\newcommand{\anticomm}[2]{\{#1,#2\}}

\begin{document}

\title{Generalized Fock--Lorentz Transformations from Projective Conformal Coordinates: Covariant Structure, Sector Classification, and Oscillator Limits}

\author{A. Boumali\,\orcidlink{0000-0002-5142-1695}}
\email{abdelmalek.boumali@univ-tebessa.dz}
\affiliation{Laboratory of Applied Physics and Theoretical Physics, Department of Matter Sciences, Faculty of Exact Sciences and Sciences of Nature and Life, Larbi Tebessi University, Tebessa, Algeria}

\author{N. Jafari\,\orcidlink{0000-0002-0285-6107}}
\email{nosrat.jafari@fai.kz}
\affiliation{Fesenkov Astrophysical Institute, 050020 Almaty, Kazakhstan}
\affiliation{Al-Farabi Kazakh National University, Al-Farabi Avenue 71, 050040 Almaty, Kazakhstan}

\author{M. Botshekananfard\,\orcidlink{0000-0003-1676-6070}}
\email{manizheh.botshekananfard@bogazici.edu.tr}
\affiliation{Department of Physics, Bo\u{g}azi\c{c}i University, 34342 Bebek, Istanbul, Turkey}

\date{\today}

\begin{abstract}
We develop a covariant projective formulation of generalized Fock--Lorentz (GFL) transformations based on the auxiliary Minkowski coordinates $X^{\mu}=x^{\mu}/[1+a_{\nu}x^{\nu}/R]$, where $R$ is a deformation length and $a^{\mu}$ is a constant deformation vector. Ordinary Lorentz transformations acting linearly on $X^{\mu}$ induce nonlinear transformations of the physical coordinates $x^{\mu}$ with a unique denominator $\mathcal{D}_{a}(x;\Lambda)$ fixed by the conformal factor. The construction gives a unified invariant interval, clarifies the singular hypersurface of the projective chart, and separates three inequivalent sectors according to the causal character of $a^{\mu}$: time-like, space-like, and null. We emphasize two points that are often obscured by analogy with the standard FL case: the coordinate velocity of light is generally defined by an implicit linear relation, and the familiar explicit FL expression is valid only in the purely time-like sector. The time-like apparent mass $m_{\rm app}(t)=m_{0}/(1+ct/R)$ and the associated one-dimensional Klein--Gordon and Dirac oscillator spectra are treated here only as limiting consistency checks of the generalized spacetime construction and are related explicitly to the companion momentum-space-dual formulation. The genuinely new dynamical result is obtained in the space-like sector, where the weak-gradient apparent mass generates a parity-breaking cubic anharmonicity; the first-order cubic shift vanishes by parity, while the combined second-order cubic and first-order quartic corrections yield a definite $R^{-2}$ shift of the oscillator operator. These results provide a transparent basis for future applications of projective relativistic kinematics without relying on a dark-universe interpretation.
\end{abstract}


\maketitle

\section{Introduction}
\label{sec:intro}

The Fock--Lorentz (FL) transformations provide a nonlinear, projective realization of Lorentz symmetry in which the usual Minkowski coordinates are replaced by fractional-linear coordinates that depend on a fundamental length scale $R$~\cite{Fock,Manida,Stepanov}. This scale is commonly interpreted as a cosmological distance, so that the FL deformation differs conceptually from the Planck-scale deformations of doubly special relativity (DSR)~\cite{Am2,Am3,MS,MS2,Kowal}. Even so, the FL framework remains a valuable setting in which Lorentz-type transformations, invariant intervals, and coordinate-dependent physical quantities can be analyzed in closed analytic form.

The standard FL transformations are usually written in terms of a single conformal factor, $1+ct/R$. In this form they are adapted to a preferred time-like direction and lead to the invariant quantity
\be
\frac{c^{2}t^{2}-\vr^{2}}{(1+ct/R)^{2}} .
\ee
A natural question is whether this construction generalizes when the time-like conformal factor is replaced by an arbitrary linear form $1+a_{\mu}x^{\mu}/R$, with $a^{\mu}$ a constant vector. Such a generalization makes it possible to distinguish the time-like, space-like, and null projective sectors, and to separate the features that are universal from those that depend on the chosen deformation direction.

The aim of this paper is to derive the generalized FL transformations within a unified and mathematically consistent notation. Four points are emphasized. First, the nonlinear transformations follow directly from ordinary Lorentz transformations acting on the auxiliary coordinates, so that no separate postulate is required. Second, the invariant interval is fixed entirely by the conformal factor. Third, the causal character of the deformation vector must be handled with care: in the metric convention $\eta_{\mu\nu}=\mathrm{diag}(-1,1,1,1)$, a null deformation vector must satisfy $a_{\mu}a^{\mu}=0$, which constrains the normalization of its spatial part. Fourth, the coordinate velocity of light is, in general, defined only implicitly when the conformal factor depends on the spatial coordinates; the familiar explicit FL expression is recovered solely in the purely time-like sector.

The present manuscript should also be distinguished from two closely related parts of our recent programme. In Ref.~\cite{BoumaliJafariDSRLF} the oscillator problem was studied in a linear--fractional DSR momentum-space setting, where the deformation acts on the Casimir invariant. In the companion FL-dual oscillator paper~\cite{BoumaliJafariBotshekananfardFLDual}, the time-like apparent mass $\mapp(t)=m_{0}/(1+ct/R)$ and its one-dimensional Klein--Gordon and Dirac oscillator spectra were derived from a momentum-space dual of the standard FL map. Here we do not present those oscillator spectra as an independent new application. Instead, we show that the same time-like apparent mass follows from a purely spacetime projective construction, thereby establishing the equivalence of the spacetime and momentum-dual routes in the time-like sector and embedding that result in a broader three-sector classification. The only oscillator detail retained beyond this cross-check is the symmetrized Klein--Gordon ordering, which replaces the unsymmetrized zero-point shift by the scalar harmonic shift $n+1/2$.

\paragraph*{Conventions.}
Throughout the paper we adopt
\be
x^{\mu}=(ct,x,y,z), \qquad \eta_{\mu\nu}=\mathrm{diag}(-1,1,1,1),
\ee
and repeated Greek indices are summed from $0$ to $3$. The covariant components of the deformation vector are $a_{\mu}=\eta_{\mu\nu}a^{\nu}$. Consistent with this mostly-plus signature, the Dirac matrices satisfy the Clifford algebra
\be
\anticomm{\gamma^{\mu}}{\gamma^{\nu}}=-2\eta^{\mu\nu},
\qquad
(\gamma^{0})^{2}=+\mathbb{1},
\qquad
(\gamma^{i})^{2}=-\mathbb{1},
\label{eq:clifford}
\ee
so that the free Dirac equation squares to the Klein--Gordon equation with the physically correct mass term.

\section{Projective rigidity and operational interpretation}
\label{sec:rigidity-operational}

The projective character of the FL map has an immediate geometric consequence: it preserves straight world lines in the auxiliary variables, but not ordinary Euclidean rigidity in the physical coordinates. Inertial motion is therefore still represented by straight trajectories, whereas the coordinate separation between initially parallel world lines may acquire a time- or position-dependent scale factor. This distinction is important, because the deformation must not be interpreted as an ordinary rigid Lorentz transformation expressed in unusual variables; rather, it is a linear-fractional realization of Lorentz symmetry in which the conformal denominator forms an intrinsic part of the physical coordinate chart.

The same observation clarifies the relation to varying-speed-of-light interpretations. Although the coordinate speed of light may depend on $t$ and $\vr$, local rods and clocks are rescaled by the same projective conformal factor. Consequently, any strictly local measurement still returns the invariant value $c$. Observable effects, should they exist, must therefore be sought in nonlocal comparisons, accumulated phases, and dimensionless ratios, or in the modified wave equations obtained after introducing a momentum-space dual. For this reason, the distinction between coordinate-dependent velocities and locally measured invariant velocities is maintained throughout the present analysis.

\section{Standard Fock--Lorentz transformations}
\label{sec:standardFL}

For a boost along the $x$ direction, with $\beta=v/c$ and $\gamma=(1-\beta^{2})^{-1/2}$, the standard FL transformations read
\bea
 t' &=& \frac{\gamma(t-vx/c^{2})}{\D_{T}(t,x;v)}, \label{eq:FL-t}\\[3pt]
 x' &=& \frac{\gamma(x-vt)}{\D_{T}(t,x;v)}, \label{eq:FL-x}\\[3pt]
 y' &=& \frac{y}{\D_{T}(t,x;v)}, \qquad
 z' = \frac{z}{\D_{T}(t,x;v)}, \label{eq:FL-yz}
\eea
where the denominator is
\be
\D_{T}(t,x;v)=1-(\gamma-1)\frac{ct}{R}+\gamma\frac{vx}{cR}.
\label{eq:DT}
\ee
As $R\to\infty$, one has $\D_{T}\to1$, and Eqs.~(\ref{eq:FL-t})--(\ref{eq:FL-yz}) reduce to the ordinary Lorentz boost. Our convention corresponds to the conformal factor $\Omega=1+ct/R$; some presentations of the FL transformations use the opposite sign of $R$, and the formulas are then mapped into the present convention by $R\mapsto -R$.

The FL transformations preserve the projective interval
\be
\frac{c^{2}t^{2}-\vr^{2}}{(1+ct/R)^{2}}=\mathrm{invariant}.
\label{eq:FL-invariant}
\ee
The structure of this invariant is most transparent in the auxiliary coordinates
\bea
T &=& \frac{t}{1+ct/R},\label{eq:Taux}\\[3pt]
\mathbf{X} &=& \frac{\vr}{1+ct/R},\label{eq:Xaux}
\eea
in which Eq.~(\ref{eq:FL-invariant}) reduces simply to $c^{2}T^{2}-\mathbf{X}^{2}$, a quantity preserved by ordinary Lorentz transformations. The nonlinearity of Eqs.~(\ref{eq:FL-t})--(\ref{eq:FL-yz}) is thus nothing more than the expression of a linear Lorentz transformation written in the projective coordinates $(t,\vr)$.

\section{Projective generalization}
\label{sec:general}

The standard construction generalizes naturally once one defines
\be
X^{\mu}=\frac{x^{\mu}}{\Om_{a}(x)}, \qquad
\Om_{a}(x)=1+\frac{a_{\nu}x^{\nu}}{R},
\label{eq:projective-map}
\ee
where the constant deformation vector $a^{\mu}$ fixes the projective direction. The inverse map is
\be
x^{\mu}=\frac{X^{\mu}}{1-a_{\nu}X^{\nu}/R}.
\label{eq:inverse-map}
\ee
Letting an ordinary Lorentz transformation act linearly on the auxiliary coordinates,
\be
X'^{\mu}=\Lambda^{\mu}{}_{\nu}X^{\nu},
\label{eq:linear-Lorentz-X}
\ee
and combining Eqs.~(\ref{eq:projective-map})--(\ref{eq:linear-Lorentz-X}), one obtains the induced nonlinear transformation of the physical coordinates,
\be
x'^{\mu}=\frac{\Lambda^{\mu}{}_{\nu}x^{\nu}}{\D_{a}(x;\Lambda)}, \qquad
\D_{a}(x;\Lambda)=1+\frac{a_{\rho}x^{\rho}-a_{\rho}\Lambda^{\rho}{}_{\sigma}x^{\sigma}}{R}.
\label{eq:general-transform}
\ee
Equation~(\ref{eq:general-transform}) is the central result of the generalized construction. It demonstrates that the denominator is not arbitrary, but is determined completely by the conformal factor of Eq.~(\ref{eq:projective-map}).

The corresponding invariant interval follows from $\eta_{\mu\nu}X^{\mu}X^{\nu}=\eta_{\mu\nu}x^{\mu}x^{\nu}/\Om_{a}^{2}(x)$ together with the Lorentz invariance of $\eta_{\mu\nu}X^{\mu}X^{\nu}$:
\be
\frac{\eta_{\mu\nu}x^{\mu}x^{\nu}}{\left(1+a_{\rho}x^{\rho}/R\right)^{2}}
=\frac{-c^{2}t^{2}+\vr^{2}}{\left(1+a_{\rho}x^{\rho}/R\right)^{2}}
=\mathrm{invariant}.
\label{eq:general-invariant-sign}
\ee
Multiplying by $-1$ gives the positive-time form
\be
\frac{c^{2}t^{2}-\vr^{2}}{\left(1+a_{\rho}x^{\rho}/R\right)^{2}}=\mathrm{invariant},
\label{eq:general-invariant}
\ee
which reduces to Eq.~(\ref{eq:FL-invariant}) for the standard time-like choice discussed below.

\paragraph*{Coordinate dependence of the conformal factor.}
Because the auxiliary coordinates are $X^{\mu}=x^{\mu}/\Om_{a}(x)$ with the single scalar denominator $\Om_{a}(x)=1+a_{\mu}x^{\mu}/R$, the way in which the transformed coordinates depend on the spatial position $x^{i}$ is determined entirely by the causal character of the constant vector $a^{\mu}$. Writing $a_{\mu}x^{\mu}=a_{0}\,ct+\mathbf{a}\cdot\vr$ in the convention $a_{\mu}=(a_{0},\mathbf{a})$, only the projection of $x^{\mu}$ onto $a^{\mu}$ enters $\Om_{a}$, and the level sets $a_{\mu}x^{\mu}=\mathrm{const}$ are the hyperplanes Minkowski-orthogonal to $a^{\mu}$. Three cases follow:
\begin{itemize}[leftmargin=1.4em,itemsep=2pt,topsep=3pt]
\item \emph{Time-like}, $a^{\mu}=(-1,\mathbf{0})$: here $a_{\mu}x^{\mu}=ct$, so $\Om_{a}=1+ct/R$ is independent of $x^{i}$. The auxiliary coordinates of Eqs.~(\ref{eq:Taux})--(\ref{eq:Xaux}), $T=t/\Om_{a}$ and $\mathbf{X}=\vr/\Om_{a}$, are rescaled isotropically by a purely time-dependent factor; the deformation is spatially homogeneous and represents a cosmological scaling common to all events on a constant-time slice.
\item \emph{Space-like}, $a^{\mu}=(0,\vn)$ with $|\vn|=1$: here $a_{\mu}x^{\mu}=\vn\cdot\vr$, so $\Om_{a}=1+\vn\cdot\vr/R$ varies linearly with the coordinate measured along $\vn$ and is independent of $t$. The transformed coordinates now depend on position through $\vn\cdot\vr$; the surfaces of constant scaling are the planes $\vn\cdot\vr=\mathrm{const}$, and the deformation is static but anisotropic, singling out a preferred spatial axis.
\item \emph{Null}, $a^{\mu}=(-1,\vn)$ with $|\vn|=1$: here $a_{\mu}x^{\mu}=ct+\vn\cdot\vr$, so $\Om_{a}=1+(ct+\vn\cdot\vr)/R$ depends on the light-front combination $u=ct+\vn\cdot\vr$ and is constant on the null hyperplanes $u=\mathrm{const}$.
\end{itemize}
Physically, the deformation vector $a^{\mu}$ selects a preferred direction in spacetime, and the dimensionless quantity $a_{\mu}x^{\mu}/R$ measures the affine displacement of the event $x^{\mu}$ from the reference hyperplane $a_{\mu}x^{\mu}=0$ along that direction. The dependence of the transformed coordinates on $x^{i}$ is therefore absent in the time-like sector (a homogeneous cosmological drift), linear along $\vn$ in the space-like sector (a preferred spatial axis), and carried by the light-front coordinate in the null sector (a preferred null hypersurface). This is the geometric origin of the homogeneous-versus-anisotropic distinction emphasized in Secs.~\ref{sec:phenomenology} and~\ref{sec:limit}.

\section{Time-like, space-like, and null sectors}
\label{sec:sectors}

The physical content of the generalized transformations is governed by the causal character of $a^{\mu}$. Since $\eta_{\mu\nu}=\mathrm{diag}(-1,1,1,1)$,
\be
a_{\mu}a^{\mu}<0 \quad \mathrm{(time\text{-}like)}, \qquad
a_{\mu}a^{\mu}>0 \quad \mathrm{(space\text{-}like)}, \qquad
a_{\mu}a^{\mu}=0 \quad \mathrm{(null)}.
\ee
Below we present the simplest representative of each class. For the space-like and null cases we employ a unit spatial vector $\vn$ with $|\vn|=1$; a symmetric direction such as $(1,1,1)$ must accordingly be written as $\vn=(1,1,1)/\sqrt{3}$ in order to preserve the normalization.

\subsection{Time-like sector: standard FL transformations}
\label{subsec:timelike}

Choosing
\be
a^{\mu}=(-1,0,0,0), \qquad a_{\mu}=(1,0,0,0), \qquad a_{\mu}x^{\mu}=ct,
\ee
and taking a boost along the $x$ direction, Eq.~(\ref{eq:general-transform}) reproduces exactly Eqs.~(\ref{eq:FL-t})--(\ref{eq:FL-yz}) with the denominator $\D_{T}$ of Eq.~(\ref{eq:DT}), and the invariant interval is Eq.~(\ref{eq:FL-invariant}).

This is the conventional FL realization. It selects a preferred time-like direction and is therefore naturally associated with a cosmological frame. The transformation is projective in spacetime, yet it reduces to ordinary special relativity as $R\to\infty$.

\subsection{Space-like sector}
\label{subsec:spacelike}

A normalized space-like deformation is
\be
a^{\mu}=(0,\vn), \qquad |\vn|=1, \qquad a_{\mu}x^{\mu}=\vn\cdot\vr.
\ee
For a boost along the same spatial direction $\vn$, set $\xi=\vn\cdot\vr$. The induced transformations are
\bea
 t' &=& \frac{\gamma(t-v\xi/c^{2})}{\D_{S}(t,\xi;v)},\label{eq:S-t}\\[3pt]
 \xi' &=& \frac{\gamma(\xi-vt)}{\D_{S}(t,\xi;v)},\label{eq:S-xi}\\[3pt]
 \vr'_{\perp} &=& \frac{\vr_{\perp}}{\D_{S}(t,\xi;v)},\label{eq:S-perp}
\eea
where $\vr_{\perp}=\vr-\xi\vn$ and
\be
\D_{S}(t,\xi;v)=1-(\gamma-1)\frac{\xi}{R}+\gamma\frac{vt}{R}.
\label{eq:DS}
\ee
For the special choice $\vn=(1,0,0)$, so that $\xi=x$, Eqs.~(\ref{eq:S-t})--(\ref{eq:S-perp}) become
\bea
 t' &=& \frac{\gamma(t-vx/c^{2})}{1-(\gamma-1)x/R+\gamma vt/R},\label{eq:Sx-t}\\[3pt]
 x' &=& \frac{\gamma(x-vt)}{1-(\gamma-1)x/R+\gamma vt/R},\label{eq:Sx-x}\\[3pt]
 y' &=& \frac{y}{1-(\gamma-1)x/R+\gamma vt/R}, \qquad
 z' = \frac{z}{1-(\gamma-1)x/R+\gamma vt/R}.\label{eq:Sx-yz}
\eea
The invariant interval is
\be
\frac{c^{2}t^{2}-\vr^{2}}{\left(1+\vn\cdot\vr/R\right)^{2}}=\mathrm{invariant}.
\label{eq:S-invariant}
\ee
In the space-like sector the conformal factor is thus controlled by the coordinate measured along the preferred spatial direction. This sector is inequivalent to the time-like FL case, as it singles out a spatial axis rather than a cosmological time direction.

\subsection{Null sector}
\label{subsec:null}

A normalized null deformation is
\be
a^{\mu}=(-1,\vn), \qquad |\vn|=1, \qquad a_{\mu}x^{\mu}=ct+\vn\cdot\vr,
\ee
which indeed satisfies $a_{\mu}a^{\mu}=-1+|\vn|^{2}=0$. For a boost along $\vn$ and $\xi=\vn\cdot\vr$, Eq.~(\ref{eq:general-transform}) yields
\bea
 t' &=& \frac{\gamma(t-v\xi/c^{2})}{\D_{N}(t,\xi;v)},\label{eq:N-t}\\[3pt]
 \xi' &=& \frac{\gamma(\xi-vt)}{\D_{N}(t,\xi;v)},\label{eq:N-xi}\\[3pt]
 \vr'_{\perp} &=& \frac{\vr_{\perp}}{\D_{N}(t,\xi;v)},\label{eq:N-perp}
\eea
with
\be
\D_{N}(t,\xi;v)=1+\left[1-\gamma(1-\beta)\right]\frac{ct+\xi}{R},
\qquad \beta=\frac{v}{c}.
\label{eq:DN}
\ee
Since $\gamma(1-\beta)=\gamma-\gamma v/c$, the coefficient in Eq.~(\ref{eq:DN}) may equivalently be written as $1-\gamma+\gamma v/c$. For $\vn=(1,0,0)$ the invariant interval is
\be
\frac{c^{2}t^{2}-\vr^{2}}{\left[1+(ct+x)/R\right]^{2}}=\mathrm{invariant},
\label{eq:N-invariant-x}
\ee
and for a general null direction $x$ is replaced by $\vn\cdot\vr$:
\be
\frac{c^{2}t^{2}-\vr^{2}}{\left[1+(ct+\vn\cdot\vr)/R\right]^{2}}=\mathrm{invariant}.
\label{eq:N-invariant}
\ee
The null sector is therefore governed by a conformal factor that depends on a light-front coordinate, and it is structurally distinct from both the time-like and the space-like sectors.

\section{Coordinate velocity of light}
\label{sec:velocity-light}

The auxiliary variables $X^{\mu}$ are Minkowskian, so that in these variables a light ray obeys
\be
\frac{d\mathbf{X}}{dT}=c\vn, \qquad |\vn|=1.
\label{eq:light-X}
\ee
In the physical coordinates $x^{\mu}$, by contrast, the velocity $\vu=d\vr/dt$ is in general coordinate dependent. Differentiating Eq.~(\ref{eq:projective-map}) yields the implicit relation
\be
\Om_{a}(x)\,\bigl(\vu-c\vn\bigr)
=\frac{\vr-ct\vn}{R}\left(ca_{0}+\mathbf{a}\cdot\vu\right),
\label{eq:implicit-speed}
\ee
where $a_{\mu}=(a_{0},\mathbf{a})$ are the covariant components entering $a_{\mu}x^{\mu}$. Equation~(\ref{eq:implicit-speed}) shows that an explicit formula for the coordinate velocity is available immediately only when the conformal factor is purely time dependent.

For the standard time-like FL case, $a_{0}=1$ and $\mathbf{a}=0$, so that Eq.~(\ref{eq:implicit-speed}) gives
\be
\vu(t,\vr;\vn)=\frac{c\vn+c\vr/R}{1+ct/R}.
\label{eq:timelike-speed}
\ee
This is the familiar FL coordinate-dependent velocity of light. In the space-like and null sectors, by contrast, $\mathbf{a}\cdot\vu$ appears on the right-hand side of Eq.~(\ref{eq:implicit-speed}), so that the velocity must be obtained by solving a linear vector equation. This is an essential caveat: substituting $\vn\cdot\vr$ or $ct+\vn\cdot\vr$ for $ct$ in Eq.~(\ref{eq:timelike-speed}) does not, in general, yield the correct coordinate velocity.

\section{Momentum-space dual and deformed wave equations}
\label{sec:momentum-dual}

For the time-like FL sector, the momentum-space dual is obtained by assigning the same projective conformal factor to the mass shell. In the generalized notation, this yields the invariant form
\be
\left(E^{2}-\vp^{2}c^{2}\right)
\left(1+\frac{a_{\mu}x^{\mu}}{R}\right)^{2}
=m_{0}^{2}c^{4},
\label{eq:general-dual-casimir}
\ee
or equivalently
\be
E^{2}-\vp^{2}c^{2}=\mapp^{2}(x)c^{4},
\qquad
\mapp(x)=\frac{m_{0}}{1+a_{\mu}x^{\mu}/R}.
\label{eq:general-mapp}
\ee
Equation~(\ref{eq:general-mapp}) defines the apparent mass. In the time-like sector it gives $\mapp(t)=m_{0}/(1+ct/R)$; in a one-dimensional space-like sector with $a^{\mu}=(0,1,0,0)$ it gives $\mapp(x)=m_{0}/(1+x/R)$; and in a null sector with $a^{\mu}=(-1,\vn)$, $|\vn|=1$, it gives $\mapp(t,\vr)=m_{0}/[1+(ct+\vn\cdot\vr)/R]$.

\paragraph*{Interpretation of the apparent mass.}
Two complementary readings of $\mapp(x)$ should be kept distinct. On the one hand, $\mapp(x)$ is, by construction, a parametrization of the projective conformal factor of Eq.~(\ref{eq:general-dual-casimir}): it absorbs the rescaling $\Om_{a}^{-2}(x)$ of the mass shell into a single position-dependent quantity, $\mapp(x)=m_{0}/\Om_{a}(x)$, while the underlying invariant mass parameter remains $m_{0}$. In this sense, $\mapp$ is a chart-dependent bookkeeping quantity rather than a new dynamical constant, and it is always traceable to the same conformal factor that defines the projective map. On the other hand, $\mapp(x)$ is the effective mass that an observer working in the physical coordinates $x^{\mu}$ would extract from the local mass-shell relation $E^{2}-\vp^{2}c^{2}=\mapp^{2}(x)c^{4}$; it is this quantity that sets the instantaneous frequency, the apparent rest energy $\mapp c^{2}$, and the oscillator level spacings derived below. The two readings are mutually consistent on account of the projective rigidity discussed in Sec.~\ref{sec:rigidity-operational}: local rods and clocks are rescaled by the same factor $\Om_{a}(x)$, so that any strictly local measurement of inertia returns the invariant value $m_{0}$, exactly as a local measurement of the velocity of light returns $c$. The variation encoded in $\mapp(x)$ is therefore observable only through nonlocal comparisons---accumulated phases, dimensionless spectral ratios, or the explicit corrections to the deformed wave equations---and never through a single local measurement. In the applications below we adopt the effective-mass language, with the understanding that $\mapp$ is in all cases nothing other than $m_{0}$ divided by the conformal factor of Eq.~(\ref{eq:general-dual-casimir}).

Promoting $E$ and $\vp$ to the standard differential operators, and treating the conformal factor as a prescribed scalar background, one obtains the generalized Klein--Gordon equation
\be
\left[
\frac{1}{c^{2}}\frac{\partial^{2}}{\partial t^{2}}
-\nabla^{2}
+\frac{m_{0}^{2}c^{2}}{\hbar^{2}\left(1+a_{\mu}x^{\mu}/R\right)^{2}}
\right]\Phi(t,\vr)=0 .
\label{eq:general-KG-equation-corrected}
\ee
For a purely time-like deformation and a single spatial momentum mode, Eq.~(\ref{eq:general-KG-equation-corrected}) reduces to
\be
\ddot{\phi}_{\mathbf{k}}(t)+
\left[
c^{2}k^{2}
+\frac{m_{0}^{2}c^{4}}{\hbar^{2}(1+ct/R)^{2}}
\right]\phi_{\mathbf{k}}(t)=0 .
\label{eq:timelike-mode-corrected}
\ee
The adiabatic parameter in the massive, long-wavelength sector ($k\to0$) is the rate of change of the instantaneous frequency $\Omega_{k}(t)$ relative to its square,
\be
\eta(t)=\left|\frac{\dot{\Omega}_{k}}{\Omega_{k}^{2}}\right|
=
\frac{\hbar}{m_{0}cR},
\label{eq:adiabatic-eta-corrected}
\ee
for $k=0$ and $1+ct/R>0$. For nonzero $k$ this expression receives the expected momentum-dependent suppression factors. The time-like mass drift is therefore adiabatically negligible whenever $R$ is of cosmological magnitude.

The corresponding Dirac equation is obtained by replacing the constant mass with the scalar field of Eq.~(\ref{eq:general-mapp}),
\be
\left(i\hbar\gamma^{\mu}\partial_{\mu}-\mapp(x)c\right)\Psi(t,\vr)=0 .
\label{eq:general-Dirac-corrected}
\ee
Squaring this first-order operator with the Clifford algebra~(\ref{eq:clifford}) generates an additional first-order term proportional to the gradient of the apparent mass,
\be
\left[\hbar^{2}\,\partial^{\mu}\partial_{\mu}
-\mapp^{2}(x)c^{2}
-i\hbar c\,\gamma^{\mu}\,\partial_{\mu}\mapp(x)\right]\Psi(t,\vr)=0 .
\label{eq:squared-Dirac-corrected}
\ee
Here $\partial^{\mu}\partial_{\mu}=-c^{-2}\partial_{t}^{2}+\nabla^{2}$, so that for constant $\mapp$ the gradient term vanishes and Eq.~(\ref{eq:squared-Dirac-corrected}) reproduces the Klein--Gordon equation~(\ref{eq:general-KG-equation-corrected}), as required by the convention~(\ref{eq:clifford}). The gradient term is suppressed by $1/R$ in the time-like sector. In the space-like sector, by contrast, it constitutes a genuine gradient coupling and cannot be discarded without an explicit approximation. For this reason, the exact instantaneous oscillator spectra derived below are restricted to the time-like sector.

\section{Oscillator limit and space-like weak-gradient correction}
\label{sec:oscillator-applications}

The time-like apparent mass
\be
\mapp(t)=\frac{m_{0}}{1+ct/R}
\label{eq:osc-mapp}
\ee
and the associated one-dimensional Klein--Gordon and Dirac oscillator spectra were derived in Ref.~\cite{BoumaliJafariBotshekananfardFLDual} from a momentum-space dual of the standard FL maps. The purpose of the present section is narrower. We show that the same mass drift follows from the purely spacetime projective construction of Sec.~\ref{sec:general}, and we use it as a consistency check of the time-like sector. The detailed Dirac matrix algebra, component reduction, and adiabatic discussion are therefore not repeated here; they are available in Ref.~\cite{BoumaliJafariBotshekananfardFLDual}. What is kept explicitly is the part that differs from the companion paper: the symmetrized Klein--Gordon ordering and the new space-like weak-gradient perturbation.

In the time-like sector the deformed Casimir may be written as
\be
\left(E^{2}-p^{2}c^{2}\right)\left(1+\frac{ct}{R}\right)^{2}=m_{0}^{2}c^{4},
\label{eq:osc-Casimir}
\ee
or equivalently $E^{2}-p^{2}c^{2}=\mapp^{2}(t)c^{4}$. For $|t|\ll R/c$,
\be
\mapp(t)=m_{0}\left(1-\frac{ct}{R}+\mathcal{O}\!\left(\frac{c^{2}t^{2}}{R^{2}}\right)\right),
\label{eq:osc-mapp-expansion}
\ee
and the instantaneous approximation is controlled by
\be
\epsilon=\frac{c}{R\omega}\ll1.
\label{eq:osc-epsilon}
\ee
For microscopic oscillator frequencies and cosmological $R$, this parameter is extremely small. Non-adiabatic phases may be treated, if needed, by the Lewis--Riesenfeld invariant or standard adiabatic-phase methods~\cite{Lewis,Berry}.

\subsection{Time-like consistency check: symmetrized KG and Dirac spectra}
\label{subsec:KGDirac-crosscheck}

For the Klein--Gordon oscillator, the ordering choice matters. Using the symmetrized Hermitian product
\be
\mathcal{O}_{\mathrm{KG}}(t)=\frac{1}{2}\left[
\left(p_{x}-i\mapp(t)\omega x\right)
\left(p_{x}+i\mapp(t)\omega x\right)
+
\left(p_{x}+i\mapp(t)\omega x\right)
\left(p_{x}-i\mapp(t)\omega x\right)
\right],
\label{eq:app-KG-hermitian}
\ee
one obtains
\be
\mathcal{O}_{\mathrm{KG}}(t)=p_{x}^{2}+\mapp^{2}(t)\omega^{2}x^{2}.
\ee
Thus the ordering-dependent constant present in an unsymmetrized product is cancelled. At fixed time the spatial eigenvalues are $\mapp(t)\hbar\omega(2n+1)$, and the leading adiabatic KG spectrum is
\be
E_{n,\mathrm{KG}}^{2}(t)=
\frac{m_{0}^{2}c^{4}}{(1+ct/R)^{2}}
+\frac{2m_{0}c^{2}\hbar\omega}{1+ct/R}\left(n+\frac{1}{2}\right),
\qquad n=0,1,2,\ldots .
\label{eq:app-KG-spectrum}
\ee
This $n+1/2$ shift is the symmetrized scalar oscillator result. A one-sided Bruce--Minning product instead leaves an ordering constant and leads to a different zero-point convention, commonly written with an $n+1$-type shift. This distinction should be stated explicitly because it is the main technical difference between the present consistency check and the companion oscillator treatment.

For the Dirac oscillator, the Moshinsky--Szczepaniak substitution $p_{x}\to p_{x}-i\mapp(t)\omega x\beta$ gives the component spectra
\bea
E_{n,+}^{2}(t)&=&\mapp^{2}(t)c^{4}+2\mapp(t)c^{2}\hbar\omega\,n,\label{eq:app-DO-upper}\\[3pt]
E_{n,-}^{2}(t)&=&\mapp^{2}(t)c^{4}+2\mapp(t)c^{2}\hbar\omega\,(n+1),\label{eq:app-DO-lower}
\eea
or, equivalently,
\be
E_{n,\pm}^{2}(t)=
\frac{m_{0}^{2}c^{4}}{(1+ct/R)^{2}}
+
\frac{2m_{0}c^{2}\hbar\omega}{1+ct/R}\,n_{\pm},
\qquad n_{+}=n,
\qquad n_{-}=n+1 .
\label{eq:app-DO-spectrum}
\ee
The two component ladders do not represent two independent physical spectra; they encode the usual pairing of upper and lower oscillator indices in the Dirac spinor. This result coincides with the time-like FL-dual result of Ref.~\cite{BoumaliJafariBotshekananfardFLDual}, as it must, because both derivations use the same conformal factor on the same world line.

\begin{figure}[H]
\centering
\includegraphics[width=0.86\linewidth]{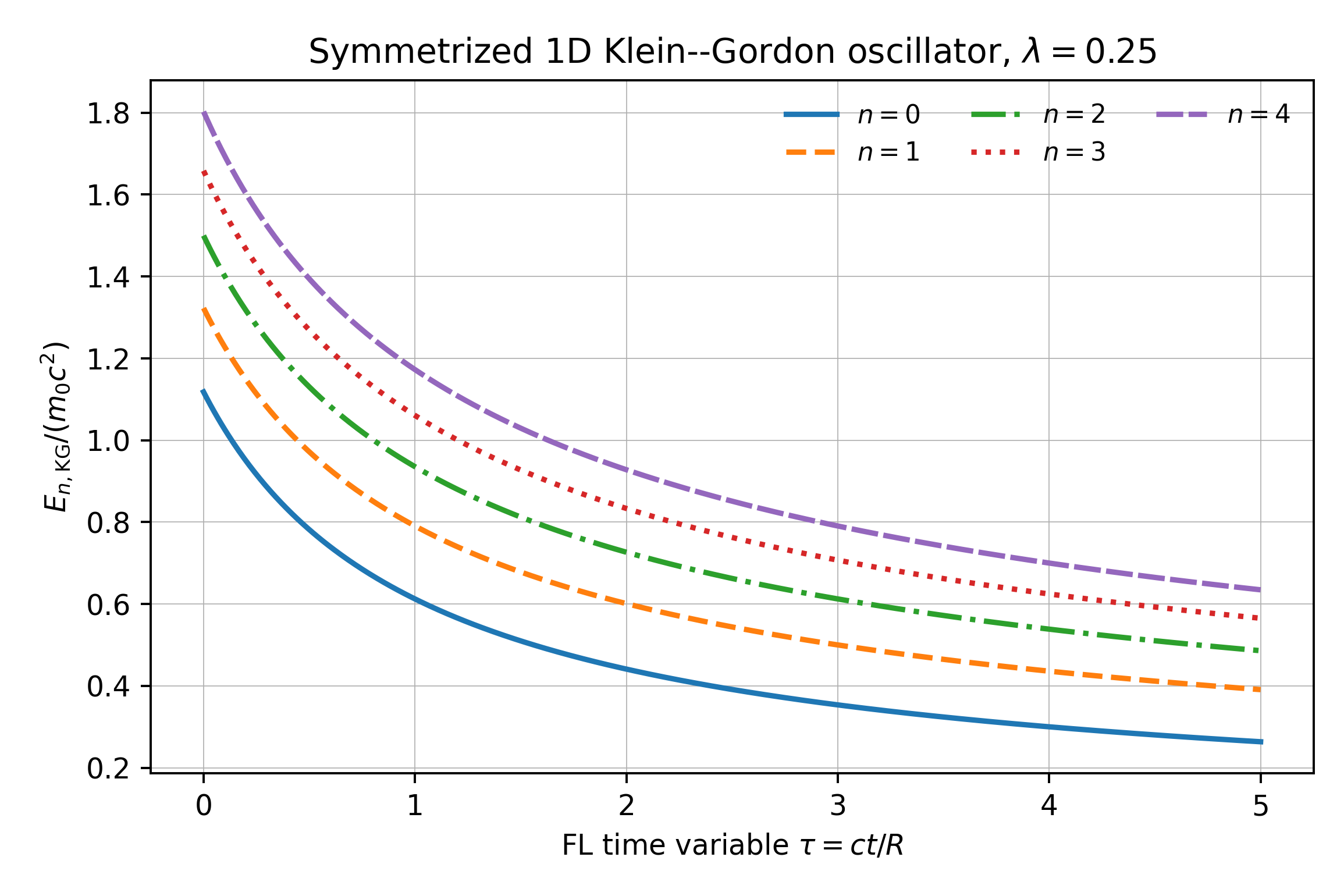}
\caption{Dimensionless positive-energy spectrum of the symmetrized one-dimensional Klein--Gordon oscillator in the time-like FL sector. The zero-point structure is $n+1/2$, consistent with Eq.~(\ref{eq:app-KG-spectrum}).}
\label{fig:kg-spectrum-time}
\end{figure}

\begin{figure}[H]
\centering
\includegraphics[width=0.86\linewidth]{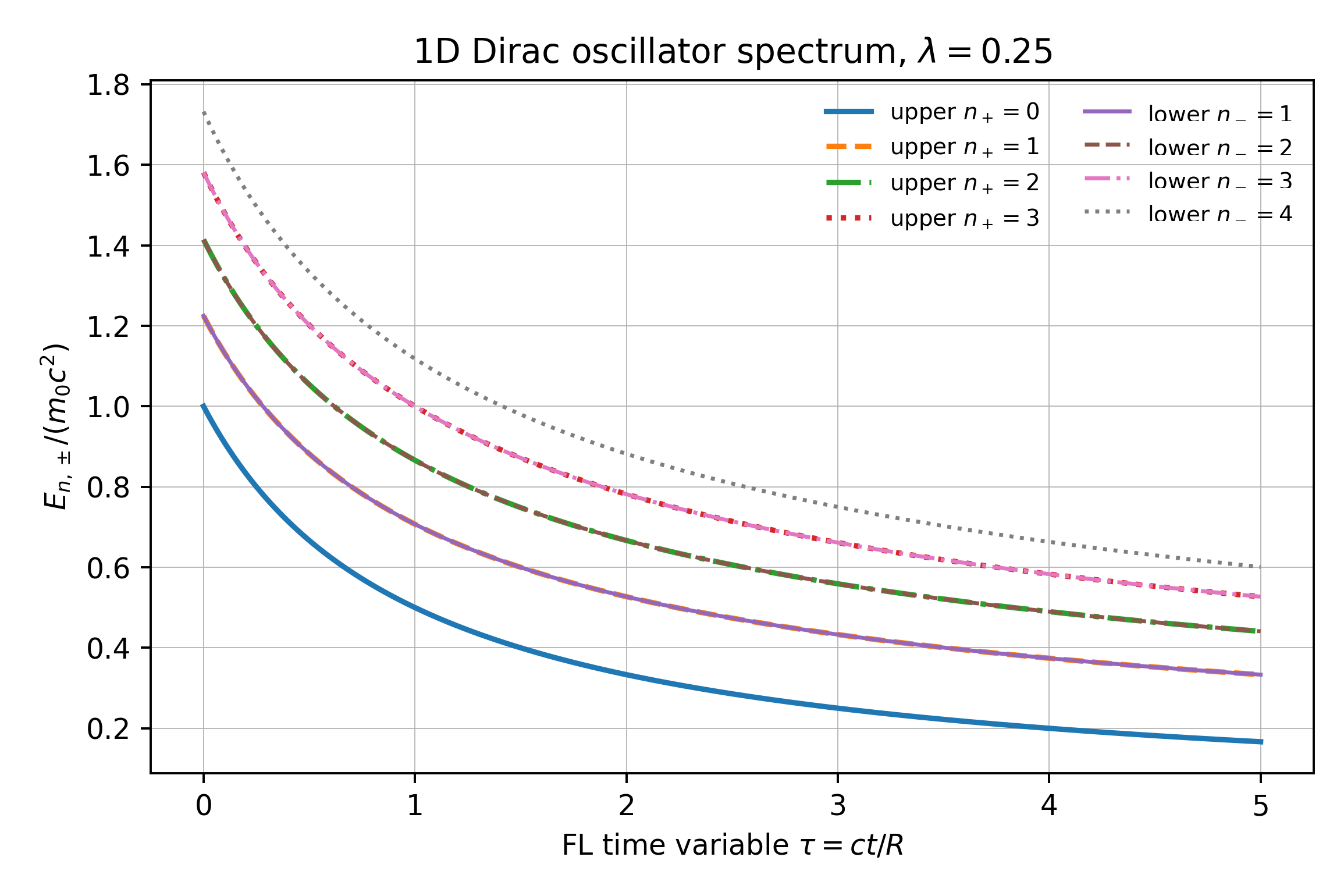}
\caption{Dimensionless positive-energy branches of the one-dimensional Dirac oscillator in the time-like FL sector for $\lambda=0.25$. The upper and lower spinor components carry the shifted indices $n_{+}=n$ and $n_{-}=n+1$.}
\label{fig:dirac-spectrum-time}
\end{figure}

\begin{figure}[H]
\centering
\includegraphics[width=0.86\linewidth]{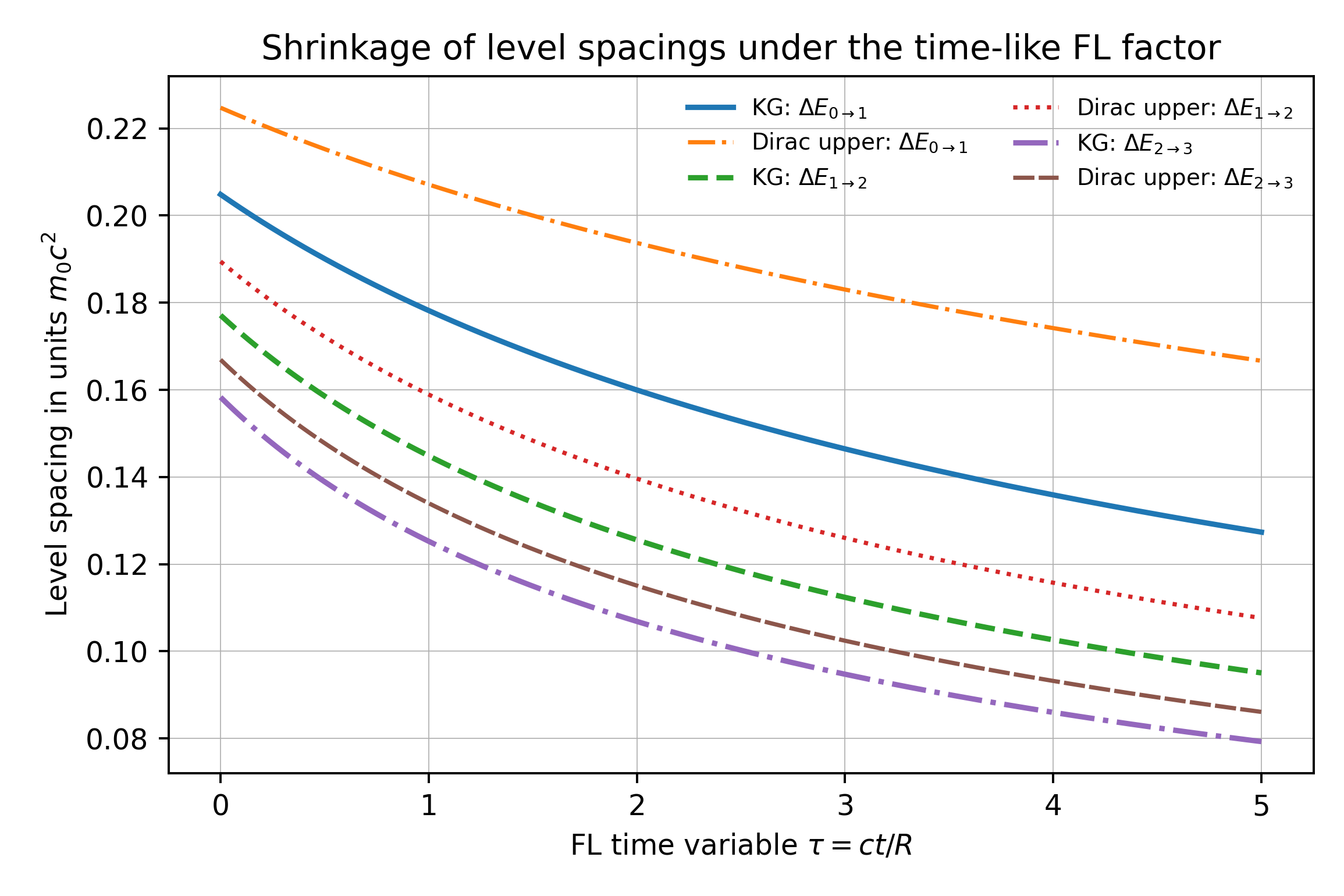}
\caption{Representative time-like level spacings. The decreasing curves simply reflect the common conformal factor $\mapp(t)=m_{0}/(1+ct/R)$ and are therefore a consistency check rather than a new independent application.}
\label{fig:kg-dirac-spacing}
\end{figure}

For visualization we use $\tau=ct/R$, $\lambda=\hbar\omega/(m_{0}c^{2})$, and $\mathcal{E}=E/(m_{0}c^{2})$. Figures~\ref{fig:kg-spectrum-time}--\ref{fig:kg-dirac-spacing} are retained as diagnostic plots, but their role is now limited to illustrating the special time-like limit of the GFL construction.

\subsection{Space-like weak-gradient oscillator: new perturbative result}
\label{subsec:spacelike-formal-osc}

The space-like sector is structurally different because the apparent mass depends on position. For $a^{\mu}=(0,1,0,0)$,
\be
\mapp(x)=\frac{m_{0}}{1+x/R},
\label{eq:space-mapp-1d}
\ee
which is singular at $x=-R$. A full quantum problem must therefore either be formulated on a domain that avoids this singular point, with suitable self-adjoint boundary conditions, or be treated as a weak-gradient effective model with $|x|\ll R$. In what follows we adopt the latter interpretation and require the oscillator length $\ell_{0}=\sqrt{\hbar/(m_{0}\omega)}$ to satisfy $\ell_{0}/R\ll1$.

The spatial part of the symmetrized oscillator operator becomes
\be
L=p_{x}^{2}+\mapp^{2}(x)\omega^{2}x^{2}.
\ee
Using
\be
\mapp^{2}(x)\omega^{2}x^{2}
=m_{0}^{2}\omega^{2}x^{2}
-2\frac{m_{0}^{2}\omega^{2}}{R}x^{3}
+3\frac{m_{0}^{2}\omega^{2}}{R^{2}}x^{4}
+\mathcal{O}\!\left(\frac{x^{5}}{R^{3}}\right),
\label{eq:space-anharmonic-expansion}
\ee
we identify a parity-breaking cubic perturbation and a quartic correction,
\be
V_{3}=-2\frac{m_{0}^{2}\omega^{2}}{R}x^{3},
\qquad
V_{4}=3\frac{m_{0}^{2}\omega^{2}}{R^{2}}x^{4}.
\ee
Let $L_{0}=p_{x}^{2}+m_{0}^{2}\omega^{2}x^{2}$, with eigenvalues
\be
\Lambda_{n}^{(0)}=m_{0}\hbar\omega(2n+1).
\ee
Since the unperturbed oscillator states have definite parity,
\be
\langle n|V_{3}|n\rangle=0.
\ee
Thus the leading effect of the cubic term appears only at second order. Standard oscillator perturbation theory gives
\be
\Delta\Lambda_{n}^{(1,4)}=
\frac{9\hbar^{2}}{4R^{2}}\left(2n^{2}+2n+1\right),
\label{eq:quartic-shift}
\ee
from the quartic term, whereas the second-order cubic contribution is
\be
\Delta\Lambda_{n}^{(2,3)}=
-\frac{\hbar^{2}}{4R^{2}}\left(30n^{2}+30n+11\right).
\label{eq:cubic-second-shift}
\ee
The combined weak-gradient correction to the spatial oscillator eigenvalue is therefore
\be
\Delta\Lambda_{n}
=
-\frac{\hbar^{2}}{2R^{2}}\left(6n^{2}+6n+1\right)
+\mathcal{O}\!\left(R^{-3}\right).
\label{eq:spacelike-total-shift}
\ee
For the corresponding symmetrized KG model, this induces
\be
\Delta E_{n}^{2}=c^{2}\Delta\Lambda_{n}
=
-\frac{\hbar^{2}c^{2}}{2R^{2}}\left(6n^{2}+6n+1\right)
+\mathcal{O}\!\left(R^{-3}\right).
\label{eq:spacelike-energy-squared-shift}
\ee
This is a genuinely space-like result: it has no analogue in the purely time-like companion oscillator paper, and it arises from the projective coordinate dependence of the GFL conformal factor.

\begin{figure}[H]
\centering
\includegraphics[width=0.86\linewidth]{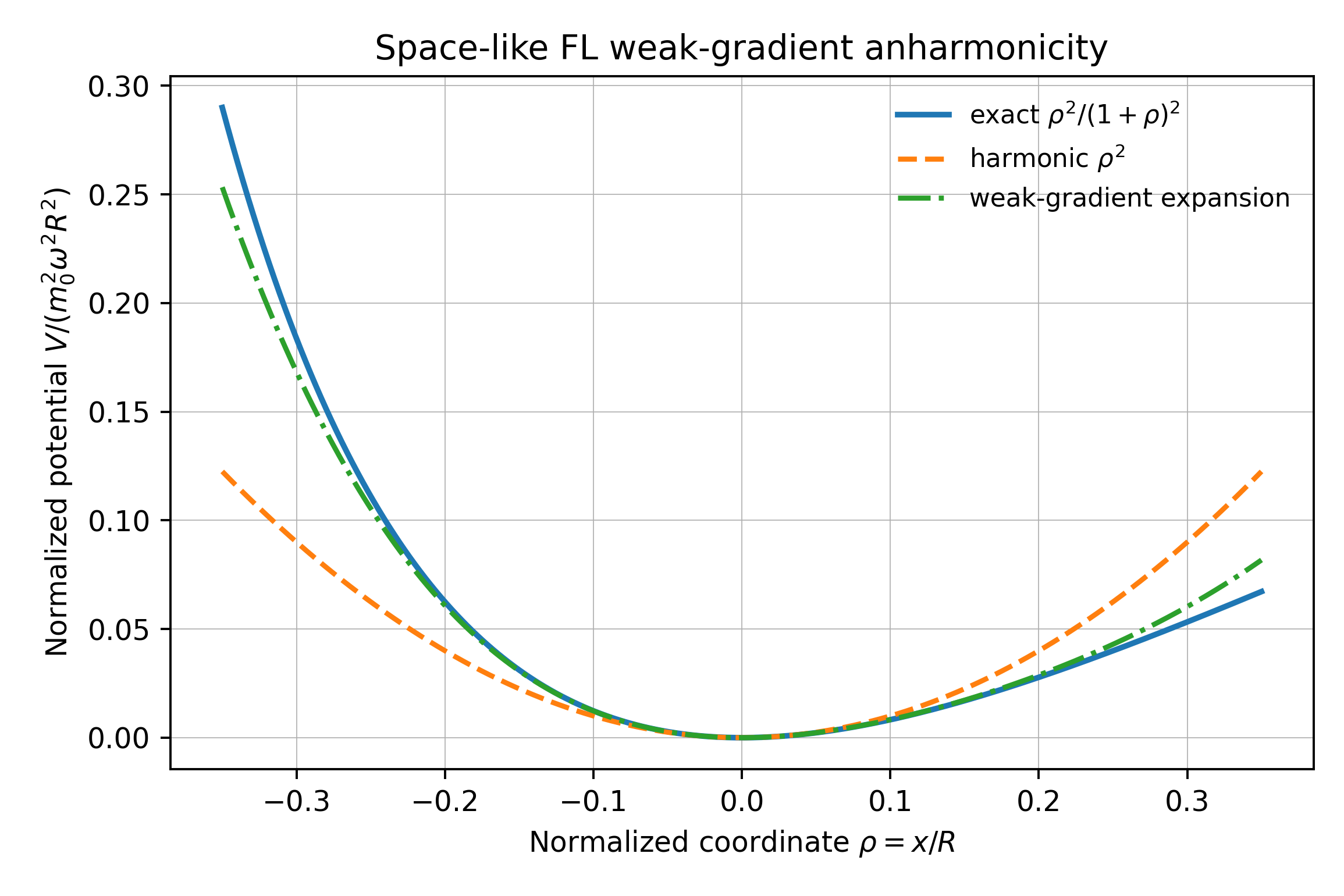}
\caption{Normalized effective oscillator potential generated by a one-dimensional space-like FL deformation. The asymmetric departure from $\rho^{2}$ illustrates the cubic anharmonicity and the associated loss of parity symmetry.}
\label{fig:space-anharmonic}
\end{figure}

\begin{figure}[H]
\centering
\includegraphics[width=0.86\linewidth]{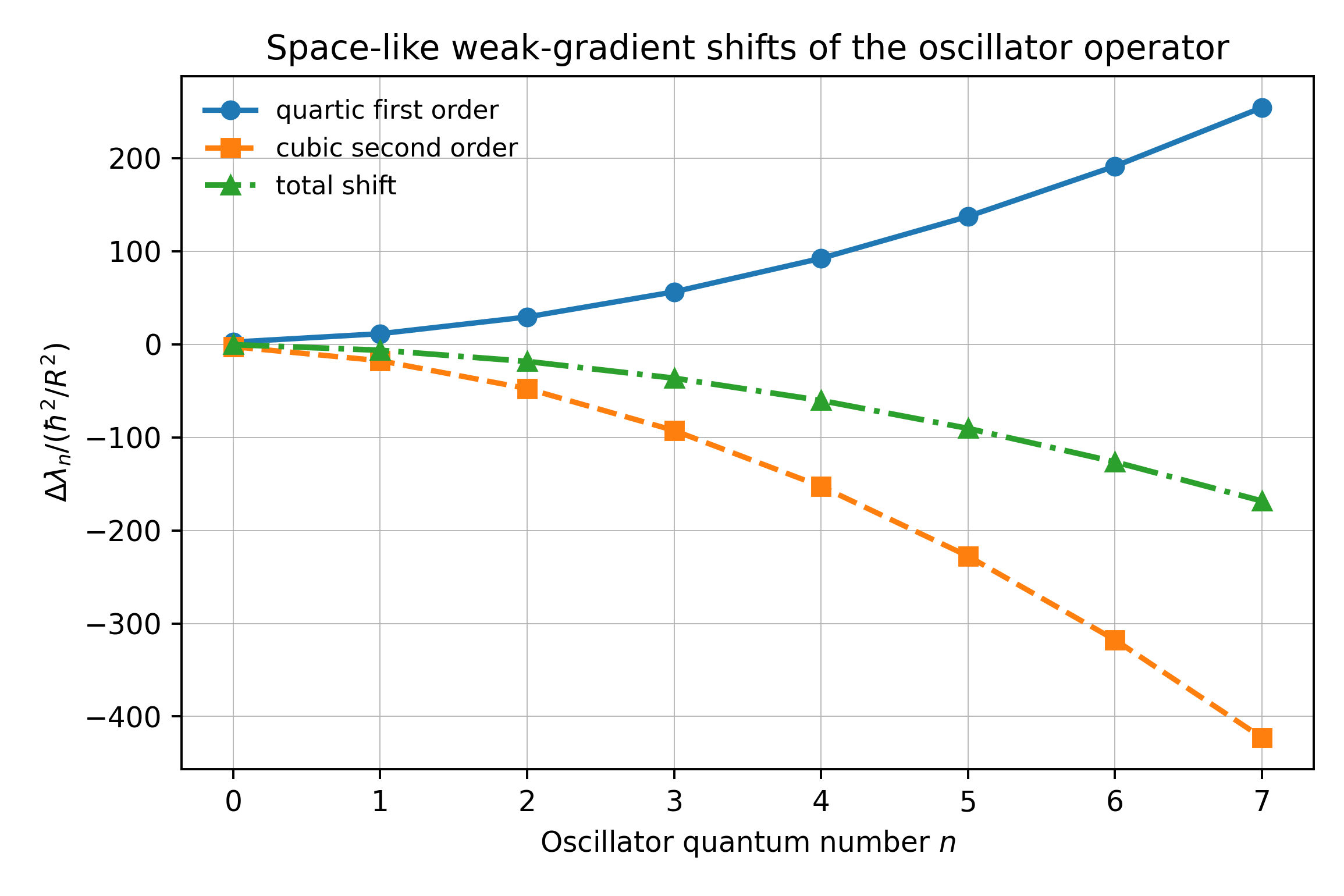}
\caption{Perturbative space-like weak-gradient shifts of the spatial oscillator operator. The first-order quartic correction is positive, the second-order cubic correction is negative, and their sum gives Eq.~(\ref{eq:spacelike-total-shift}).}
\label{fig:spacelike-shifts}
\end{figure}

Figure~\ref{fig:space-anharmonic} compares the exact normalized space-like factor $\rho^{2}/(1+\rho)^{2}$, with $\rho=x/R$, with the harmonic limit and the weak-gradient expansion. Figure~\ref{fig:spacelike-shifts} shows how the quartic first-order and cubic second-order corrections combine into the total shift of Eq.~(\ref{eq:spacelike-total-shift}).

\section{Phenomenological comments}
\label{sec:phenomenology}

The preceding formulas show that the local variation induced by a cosmological scale $R$ is extremely small. If $R$ is identified with a Hubble-scale length, the fractional time variation of the time-like mass is of order $c/R$, comparable to the Hubble rate $H_{0}$. Direct laboratory detection is therefore not expected. More realistic constraints would have to arise from dimensionless comparisons over cosmological baselines, such as high-redshift spectral ratios, accumulated phase differences, or directional tests in very-high-energy astrophysical propagation.

For the space-like and null sectors, the characteristic signature is anisotropy rather than a uniform mass drift. The preferred vector $\vn$ enters the invariant interval, the apparent mass, and the wave equation, so that any phenomenological claim based on these sectors must be framed in terms of directional observables. This conclusion mirrors the mathematical distinction drawn above: the time-like sector produces a homogeneous cosmological scaling, whereas the space-like and null sectors select preferred spatial or light-front directions.

\section{Special-relativistic limit and physical interpretation}
\label{sec:limit}

All sectors considered above share the same limiting behavior. As $R\to\infty$ the conformal factor tends to unity,
\be
\Om_{a}(x)=1+\frac{a_{\mu}x^{\mu}}{R}\longrightarrow 1,
\ee
and the generalized transformations reduce to
\be
x'^{\mu}=\Lambda^{\mu}{}_{\nu}x^{\nu},
\ee
with the invariant interval reducing to the standard Minkowski interval. The generalized FL construction is thus a deformation of special relativity controlled by the dimensionless ratio $a_{\mu}x^{\mu}/R$.

The various choices of $a^{\mu}$ should not be regarded as a single transformation expressed in different coordinates. They correspond to distinct projective structures, since they select different preferred directions in spacetime. The time-like case is naturally associated with cosmological time. The space-like case introduces a preferred spatial direction and is therefore meaningful only as an anisotropic toy model. The null case is adapted to a light-front coordinate and may prove useful for studying deformations associated with radiation-like propagation or null hypersurfaces. A complete dynamical theory would have to specify which sector is physically realized and how matter fields couple to the corresponding projective geometry.

The oscillator limit illustrates the physical role of the time-like sector without being the main novelty of the present paper. The same conformal factor that defines the projective FL map generates the apparent mass $\mapp(t)$ and therefore reproduces the time-like oscillator spectra obtained by the momentum-dual route in Ref.~\cite{BoumaliJafariBotshekananfardFLDual}. The limit $R\to\infty$ simultaneously restores the Lorentz transformations, the constant rest mass $m_{0}$, and the usual relativistic oscillator spectra. In contrast, the space-like weak-gradient calculation gives a new anisotropic correction controlled by $\ell_{0}/R$, with the domain restricted by the singular surface $x=-R$.

\section{Conclusion}
\label{sec:conclusion}

We have reformulated the generalized Fock--Lorentz transformations by means of a unified projective map, $X^{\mu}=x^{\mu}/[1+a_{\nu}x^{\nu}/R]$. Ordinary Lorentz transformations in the auxiliary Minkowski coordinates induce nonlinear transformations of the physical coordinates, with the denominator determined by Eq.~(\ref{eq:general-transform}). This derivation clarifies the origin of the invariant interval and removes the ambiguities that arise when the denominator is inferred by analogy with the standard FL case.

The standard FL transformations are recovered for a time-like deformation vector, $a^{\mu}=(-1,0,0,0)$. The space-like and null sectors are obtained by choosing $a^{\mu}=(0,\vn)$ and $a^{\mu}=(-1,\vn)$, respectively, with $|\vn|=1$. This normalization is essential: in the present metric convention, the vector $(-1,1,1,1)$ is not null unless its spatial part is appropriately normalized. The corresponding invariant intervals are controlled by $1+\vn\cdot\vr/R$ in the space-like case and by $1+(ct+\vn\cdot\vr)/R$ in the null case.

The time-like oscillator limit was reframed as a consistency check rather than as a new independent application. The apparent mass $\mapp(t)=m_{0}/(1+ct/R)$ and the associated Dirac oscillator spectrum coincide with the companion momentum-dual FL treatment, while the symmetrized Klein--Gordon ordering used here gives the scalar zero-point shift $n+1/2$. This explicit cross-reference removes any ambiguity about overlap: the repeated time-like result is used only to verify the equivalence of the spacetime-projective and momentum-dual routes.

Finally, we derived the general coordinate-velocity relation for light. The familiar explicit FL expression holds only in the purely time-like sector, whereas the space-like and null sectors require the solution of an implicit linear relation. The new quantitative space-like result is the weak-gradient oscillator shift, $\Delta\Lambda_{n}=-(\hbar^{2}/2R^{2})(6n^{2}+6n+1)+\mathcal{O}(R^{-3})$, obtained after the vanishing of the first-order cubic contribution by parity. Taken together, these results provide a cleaner basis for future applications of generalized FL transformations to relativistic wave equations, deformed dispersion relations, and anisotropic extensions of relativistic kinematics.

\section*{Acknowledgments}

The authors thank their colleagues for useful discussions on projective realizations of Lorentz symmetry and generalized relativistic transformations.

\end{document}